\documentclass[10pt,aps,pre,twocolumn,superscriptaddress,longbibliography,reprint]{revtex4-1}

\usepackage[english]{babel}
\usepackage[utf8]{inputenc}
\usepackage[colorlinks=true,citecolor=blue]{hyperref}

\usepackage{amsmath,amssymb,bbm,mathtools}
\usepackage[inline,shortlabels]{enumitem}
\usepackage[per-mode=symbol-or-fraction,retain-unity-mantissa=false]{siunitx}

\DeclareMathOperator{\trace}{tr}
\DeclareMathOperator{\artanh}{artanh}
\newcommand{\tr}[1]{\trace\bigl\{ #1\bigr\}}

\DeclarePairedDelimiter{\bra}{\langle}{\rvert}
\DeclarePairedDelimiter{\ket}{\lvert}{\rangle}
\DeclarePairedDelimiterX{\braket}[2]{\langle}{\rangle}{#1\,\delimsize\vert\,\mathopen{}#2}
\DeclarePairedDelimiter{\av}{\langle}{\rangle}
\def \llangle {\langle\!\langle}
\def \rrangle {\rangle\!\rangle}
\DeclarePairedDelimiterX{\corr}[2]{\llangle}{\rrangle}{#1 ; #2}
\DeclarePairedDelimiterX{\comm}[2]{[}{]}{#1, #2}
\DeclarePairedDelimiter{\abs}{\lvert}{\rvert}
\DeclarePairedDelimiter{\norm}{\lVert}{\rVert}

\newcommand{\intd}[3]{\int_{#1}^{\mathmakebox[0pt][l]{#2}\hphantom{\T}}\!\!\!\! d{#3} \;}
\newcommand{\intD}[3]{\int_{#1}^{#2}\!\!\!\! \mathcal D{#3} \;}
\newcommand{\tint}{\intd{0}{\T}{t}}
\newcommand{\indefInt}[1]{\int^{\hphantom{\T}}\!\!\!\! d{#1} \;}
\newcommand{\ksum}{\sum\nolimits_k\!}
\newcommand{\rsum}{\sum\nolimits_\nu\!}
\newcommand{\asum}{\sum\nolimits_\alpha\!}
\newcommand{\jsum}{\sum\nolimits_j\!}

\newcommand{\T}{\mathcal{T}}
\newcommand{\D}{\mathsf{D}}
\renewcommand{\L}{\ensuremath \mathsf{L}}

\renewcommand{\P}{\ensuremath \mathcal{P}}

\newcommand{\R}{\mathcal{R}}
\newcommand{\A}{\mathcal{A}}
\newcommand{\Xabs}{\Xi}

\newcommand{\quant}{\textnormal{c}}
\newcommand{\cool}{\textnormal{c}}
\newcommand{\ssIn}{\textnormal{in}}
\newcommand{\jump}{\textnormal{j}}
\newcommand{\qb}{\textnormal{qb}}
\newcommand{\thermal}{\textnormal{th}}
\newcommand{\tot}{\textnormal{tot}}
\newcommand{\Carnot}{\textnormal{C}}

\begin{document}

%Title of paper
\title{Quantum jump approach to microscopic heat engines}

\author{Paul Menczel}
\affiliation{Department of Applied Physics, Aalto University, 00076 Aalto, Finland}

\author{Christian Flindt}
\affiliation{Department of Applied Physics, Aalto University, 00076 Aalto, Finland}

\author{Kay Brandner}
\affiliation{School of Physics and Astronomy, University of Nottingham, Nottingham NG7 2RD, United Kingdom}
\affiliation{Centre for the Mathematics and Theoretical Physics of Quantum Non-Equilibrium Systems, University of Nottingham, Nottingham NG7 2RD, United Kingdom}

\begin{abstract}
Modern technologies could soon make it possible to investigate the operation cycles of quantum heat engines by counting the photons that are emitted and absorbed by their working systems.
Using the quantum jump approach to open-system dynamics, we show that such experiments would give access to a set of observables that determine the trade-off between power and efficiency in finite-time 
engine cycles.
By analyzing the single-jump statistics of thermodynamic fluxes such as heat and entropy production, we obtain a family of general bounds on the power of microscopic heat engines.
Our new bounds unify two earlier results and admit a transparent physical interpretation in terms of single-photon measurements. 
In addition, these bounds confirm that driving-induced coherence leads to an increase in dissipation that suppresses the efficiency of slowly driven quantum engines in the weak-coupling regime. 
A nanoscale heat engine based on a superconducting qubit serves as an experimentally relevant example and a guiding paradigm for the development of our theory.  
\end{abstract}

\maketitle

\section{Introduction}

In classical thermodynamics, a heat engine is described as a machine that uses a periodic series of thermodynamic processes to convert thermal energy into mechanical work \cite{Zemansky1997}.
Each process, or stroke, involves the transfer of work between the working medium and an external load, for example a movable piston, or the exchange of heat with either a hot or a cold reservoir.
Output and input of the engine, that is, the net generated work and the heat uptake from the hot reservoir, depend on the applied protocol and the equations of state of the medium.
Their ratio, however, the thermal efficiency, is subject to a universal upper bound, the Carnot limit, which follows directly from the first and the second law and is attained for optimal quasi-static, i.e., infinitely long, cycles \cite{Zemansky1997}.

Realistic machines have to operate at finite speed and therefore inevitably produce dissipative losses, which suppress their efficiency.
But how close can a heat engine with fixed cycle time come to the Carnot limit?
This question cannot be resolved within the framework of classical thermodynamics due to its lack of a fundamental time scale.
Early on, this issue spurred the development of refined models for macroscopic heat engines that account for irreversible effects by phenomenological means, an approach that became known as finite-time thermodynamics \cite{ChenJNon-EquilibThermodyn1999, SalamonEnergy2001, HoffmannJNon-EquilibThermodyn2003, AndresenAngewChemIntEd2011}.
More recently, with the advent of stochastic and quantum thermodynamics, the focus has changed to microscopic heat engines, which, instead of a homogeneous medium, use a tiny object with few degrees of freedom to perform thermodynamic cycles \cite{SeifertRepProgPhys2012, KosloffEntropy2013, VinjanampathyContempPhys2016, BenentiPhysRep2017, Binder2018}.
The input for such devices is provided by tunable heat sources, which control the temperature of the environment of the working system;	work is extracted and injected by changing the internal energy of the system through external control parameters or by coupling it to a microscopic load.

In contrast to a macroscopic fluid, which contains a vast amount of particles, the working systems of a microscopic heat engine can be described on the level of trajectories or wave functions.
Macroscopic equations of state are thereby replaced by stochastic equations of motion, which apply even far from equilibrium and create a direct link between micro-dynamics and thermodynamics.
This approach has opened a wide range of possibilities to explore the basic principles that govern the performance of periodic heat engines.
Recent investigations include the study of generalized cycles, which involve continuous temperature variations \cite{BrandnerPhysRevX2015, RazPhysRevLett2016, BrandnerPhysRevE2016, CerinoPhysRevE2016, BrandnerPhysRevLett2020}, the development of optimal control strategies \cite{SchmiedlEPL2007, EspositoPhysRevE2010, DechantPhysRevLett2015, BauerPhysRevE2016, CavinaPhysRevA2018, MenczelPhysRevB2019, AbiusoPhysRevA2019, ErdmanNewJPhys2019, AbiusoPhysRevLett2020}, and the systematic investigation of the thermodynamic footprint of quantum effects, which become relevant at time and energy scales comparable to Planck's constant, see for instance \cite{FunoPhysRevA2013, UzdinPhysRevX2015, UzdinPhysRevAppl2016, WatanabePhysRevLett2017, BrandnerPhysRevLett2017, FriedenbergerEPL2017, NiedenzuNatCommun2018, CamatiPhysRevA2019, GhoshEurPhysJSpecTop2019, PekolaPhysRevB2019, DannNewJPhys2020, LatuneNewJPhys2020}.
As a key result, this development led to the discovery of a family of trade-off relations between power, i.e., average work output per unit time, and efficiency, first in linear response \cite{BrandnerPhysRevX2015, BrandnerPhysRevE2016, ProesmansPhysRevLett2016} and then beyond \cite{ShiraishiPhysRevLett2016, ShiraishiJStatPhys2019, KoyukPhysRevLett2019}.
These relations impose quantitative bounds on the efficiency of finite-time engine cycles, which go beyond the first and the second law and approach the Carnot limit only for infinite cycle times leading to vanishing power.
This behavior is, in fact, generic for conventional systems and can be overcome only under exceptional conditions such as diverging fluctuations of thermodynamic quantities \cite{CampisiNatCommun2016, HolubecPhysRevE2017, KoyukPhysRevLett2019} or vanishing relaxation times enabled by fine-tuned dissipation mechanisms \cite{AllahverdyanPhysRevLett2013, PolettiniEPL2017, HolubecPhysRevE2017a, HolubecPhysRevLett2018}.

Over the last decade, microscopic heat engines have been realized on increasingly smaller length and energy scales with working systems such as a micrometer-sized silicon spring \cite{SteenekenNaturePhys2011}, colloidal particles \cite{BlickleNaturePhys2012, MartinezPhysRevLett2015, MartinezNaturePhys2016, KrishnamurthyNaturePhys2016, ProesmansPhysRevX2016}, a single atom \cite{AbahPhysRevLett2012, RossnagelScience2016}, nuclear \cite{PetersonPhysRevLett2019} and electronic \cite{vonLindenfelsPhysRevLett2019} spins or nitrogen-vacancy centers in diamond \cite{KlatzowPhysRevLett2019}.
In light of this rapid development, practical tests of trade-off relations between power and efficiency appear as a realistic challenge for near-future experiments.
This endeavor will, besides measurements of produced work and absorbed heat, require the measurement of at least one additional parameter, which is necessary to match the physical dimensions of power and efficiency \cite{PietzonkaPhysRevLett2018, KoyukPhysRevLett2019}.
Despite its practical importance, this problem has so far received only little attention and studies of such trade-off relations in the quantum regime have so far been focused on specific examples \cite{SacchiArXiv200705399Quant-Ph2020} or limiting regimes \cite{MillerArXiv200607316Quant-Ph2020}.

In this article, we consider a promising solid-state platform for the realization of quantum thermal devices that is in reach of current technologies \cite{PekolaNewJPhys2013, HekkingPhysRevLett2013, PekolaNaturePhys2015, CampisiNewJPhys2015, ViisanenNewJPhys2015, DonvilPhysRevA2018}.
This setup consists of an engineered working system and a mesoscopic reservoir, which acts as an effective environment.
An engine cycle can be implemented by applying a periodic driving field to the system and modulating the base temperature of the reservoir.
At the same time, the temperature of the reservoir is monitored with an ultra-sensitive thermometer able to detect small variations due to the emission and absorption of single photons \cite{GasparinettiPhysRevAppl2015, GoveniusPhysRevLett2016, KarimiPhysRevApplied2018, WangApplPhysLett2018}.
Each detected event indicates the transfer of a specific amount of energy between the system and the reservoir and an abrupt change of the quantum state of the system, in other words, a quantum jump.
Hence, the reservoir takes on a two-fold role; it functions as a source of thermal energy and as a small-scale calorimeter enabling the direct measurement of the exchanged heat and the quantitative observation of quantum jumps.
As we show in the following, the statistics of these jumps encodes an operationally accessible trade-off relation between power and efficiency for quantum heat engines.

The quantum jump approach to dissipative dynamics was first conceived for applications in quantum optics \cite{DalibardPhysRevLett1992, MolmerJOptSocAmB1993, PlenioRevModPhys1998, Breuer2002} and has long been recognized as a powerful tool to extend the notions of stochastic thermodynamics into the quantum regime \cite{BreuerPhysRevA2003, Binder2018}.
The quantum jump record is thereby commonly treated as an analogue of a classical trajectory, along which fluctuating thermodynamic quantities can be consistently defined by invoking the two-point measurement scheme \cite{EspositoRevModPhys2009, CampisiRevModPhys2011}. 
Here, we pursue an alternative approach: instead of considering accumulated quantities over an entire record, we focus on the statistics of the quanta of thermodynamic fluxes that are exchanged between the system and its environment in individual quantum jumps.
This strategy opens a new perspective on thermodynamic processes in systems with quantized energy levels and enables us to built a connection between the microscopic anatomy of the energy flow in a quantum engine cycle and its overall performance.

Our paper is organized as follows.
In the next section, we set the stage by briefly reviewing the basics of the experimental setup proposed in
Refs.~\cite{PekolaNewJPhys2013, HekkingPhysRevLett2013, PekolaNaturePhys2015, CampisiNewJPhys2015, ViisanenNewJPhys2015, DonvilPhysRevA2018} and discuss how it can be used to realize a microscopic heat engine
with a superconducting qubit.
We also introduce the concept of single-jump distributions and illustrate this idea with a numerical simulation.
In Sec.~\ref{Sec_GenTheor}, we set up the theoretical framework for our analysis and proceed in several steps towards our main result:
a family of new trade-off relations between power and efficiency for quantum heat engines, which hold for arbitrary multi-level systems and driving protocols.
These relations involve only physically transparent parameters that can be determined through single-photon measurements and they unify several earlier results, which we recover as special cases. 
To demonstrate the quality of our bounds, we apply them to the qubit engine of Sec.~\ref{Sec_QubitI} in Sec.~\ref{Sec_QubitII}.
We discuss possible directions for future research and conclude in Sec.~\ref{Sec_DisPers}.

\section{Qubit Engine: Setup} \label{Sec_QubitI}

\begin{figure*}
	\centering
	\includegraphics[width=\textwidth]{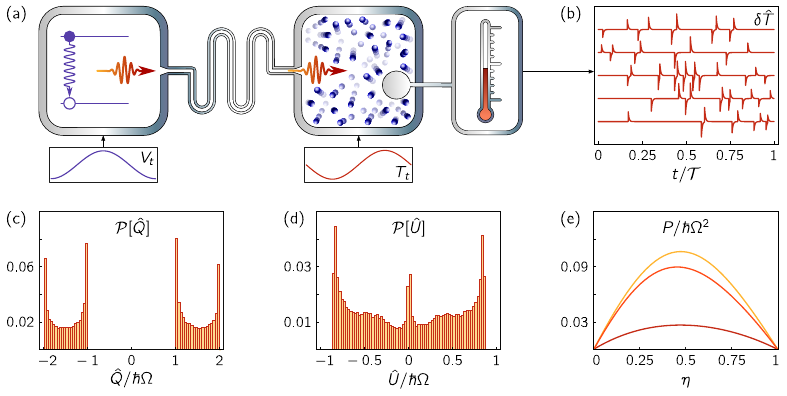}
	\caption[.]{
		Quantum jump thermodynamics of a single-qubit engine.
		\begin{enumerate*}[(a)]
		\item Setup.
			The left box represents the qubit.
			The separation between the two energy levels is sketched as horizontal lines.
			The two boxes on the right are the calorimeter, which consists of the electron gas inside a metallic island, and an ultra-sensitive thermometer monitoring its temperature.
			The small panels on the bottom show the driving protocols for the level splitting of the qubit and the base temperature of the island, cf.\ Eq.~\eqref{QT_Prot}.
			Each photon exchange between the qubit and the island is accompanied by a quantum jump of the qubit and a short-lived variation of the electron temperature.
		\item Temperature fluctuations of the electron gas, $\delta \hat{T}_t \equiv \hat{T}_t - T_t$, as detected by the thermometer for five different operation cycles.
		\item Single-jump distribution of the exchanged heat between qubit and reservoir, cf.\ Eq.~\eqref{QT_QDist}.
		\item Single-jump distribution of the effective thermal input from the heat source, cf.\ Eq.~\eqref{QT_UDist}.
		\item Power vs efficiency for $\Delta = 0,\;0.5,\;0.95$ from top to bottom, cf.\ Eq.~\eqref{QT_Eff}.
		\end{enumerate*}
		The plots in panels (b)-(d) were obtained from a stochastic wave-function simulation with the jump operators \eqref{QT2_JOp} for $\T=1/\Omega$ \cite{Breuer2002}.
		For the plots (c) and (d), averages were taken over 5000 cycles.
		The plot (e) was prepared by numerically calculating the time-dependent density matrix of the qubit for increasing cycle times running from $\T=0.1/\Omega$ to $\T=250/\Omega$.
		For all numerical calculations we have set $\kappa=10$ and $T_0=\hbar\Omega$, see Sec.~\ref{Sec_QubitII} for details.}
	\label{Fig_Setup}
\end{figure*}

We consider a solid-state realization of a microscopic heat engine with a superconducting qubit.
This working system can be described by the Hamiltonian%
\begin{equation} \label{QT_Hamiltonian}
	H^\qb = -\frac{\hbar\Omega}{2} \Bigl( \Delta \sigma_x + \sqrt{V^2-\Delta^2} \sigma_z \Bigr) ,
\end{equation}
where $\Omega$ sets the overall energy scale.
The dimensionless parameters $\Delta$ and $V$ correspond to the characteristic tunneling energy of the qubit and the level splitting, which can be controlled by varying the bias magnetic flux \cite{ChiorescuScience2003, NiskanenPhysRevB2007}; $\sigma_x$ and $\sigma_z$ are the usual Pauli matrices and $\hbar$ denotes Planck's constant.
The qubit is coupled to the electronic degrees of freedom of a metallic island, which acts as a mesoscopic reservoir, see Fig.~\ref{Fig_Setup}(a).
An engine cycle is realized by periodically changing the level separation $V$ and the base temperature $T$ of the island.
For simplicity, we here focus on harmonic driving protocols given by%
\begin{equation} \label{QT_Prot}
	V_t = \frac{1}{2} \bigl( 3 - \cos[2\pi t/\T] \bigr)
		\;\;\text{and}\;\;
	T_t = \frac{T_0}{2} \bigl( 3 - \sin[2\pi t/\T] \bigr) ,
\end{equation}
so that the dimensionless level splitting $V_t$ and the normalized temperature $T_t / T_0$ oscillate between $1$ and $2$, where $T_0$ is the base temperature of the island and $\T$ denotes the cycle time.

In the low-temperature regime, the electron gas inside the island features a low heat capacity and a short internal relaxation time of the order of nanoseconds.
After absorbing or emitting a photon, the electron gas therefore first settles to an internal equilibrium state before returning to the base temperature of the island via electron-phonon mediated heat flow to the substrate.
This equilibration process takes place on a much longer time scale, on the order of $\SI{100}{\nano\second}$ \cite{KupiainenPhysRevE2016, DonvilPhysRevA2018, GuarcelloPhysRevApplied2019}.
This mechanism leads to spikes and dips in the temperature trace $\hat{T}_t$ of the electron gas, which should not be confused with the base temperature of the island $T$.
Since the temperature of the electron gas can be detected with an ultra-sensitive electron thermometer, see Fig.~\ref{Fig_Setup}(b), it becomes possible to detect the exchange of single photons and to measure a quantum jump record,%
\begin{equation}
	\R = \bigl\{ (t_k,d_k) \bigr\} ,
\end{equation}
for every operation cycle of the engine.
Here, $t_k$ is the time at which the event $k$ occurs and the binary variable $d_k$ indicates whether a photon was emitted ($d_k=+$) or absorbed ($d_k=-$) by the reservoir, that is, whether the qubit jumped to its excited or ground state.

Each detected event indicates the transfer of discrete amounts of heat and effective thermal energy, $\hat{Q}$ and $\hat{U}$, between the qubit and the reservoir.
The statistics of these thermal quanta is described by the single-jump distributions%
\begin{subequations}
	\begin{align} \label{QT_QDist}
		\P[\hat{Q}] &= \frac{1}{\A} \mathbb{E} \Bigl[ \ksum \delta\bigl[ \hat{Q}-d_k\varepsilon_{t_k} \bigr] \Bigr]
		\quad\text{and} \\[6pt]
		\P[\hat{U}] &= \frac{1}{\A} \mathbb{E} \Bigl[ \ksum \delta\bigl[ \hat{U}-d_k\varepsilon_{t_k}\eta_{t_k} \bigr] \Bigr] , \label{QT_UDist}
	\end{align}
\end{subequations}
which can be determined by running the experiment over a large number of cycles, see Figs.~\ref{Fig_Setup}(c) and \ref{Fig_Setup}(d).
Here, the symbol $\mathbb{E}$ indicates the average over all jump records, $\varepsilon_t \equiv \hbar\Omega V_t$ is the time-dependent level splitting of the qubit, $\eta_t \equiv 1-T_0/T_t$ is the instantaneous Carnot factor and the activity $\A = \mathbb{E} \bigl[\, \ksum 1\, \bigr]$ is a normalization factor corresponding to the mean number of jumps per cycle.
Note that, throughout this article, we use hats to indicate thermodynamic quantities associated with single quantum jumps and $\delta$ denotes the Dirac delta function.

The mean value $\av{\hat{Q}} \equiv Q/\A$ of the heat flux $\hat{Q}$ determines the average output of the engine since the first law requires $Q=W$, where $Q$ is net heat uptake of the working system per cycle and $W$ the produced work.
In analogy, the mean value $\av{\hat{U}} \equiv U/\A$ of the flux $\hat{U}$ can be regarded as the effective input provided by the external heat source \cite{BrandnerPhysRevX2015, BrandnerPhysRevE2016}. 
This identification leads to a consistent generalization of the standard thermal efficiency for heat engines that are driven by continuous temperature variations,%
\begin{equation} \label{QT_Eff}
	\eta \equiv W/U  = \av{\hat{Q}}/\av{\hat{U}} \leq 1 .
\end{equation}
The upper bound $1$ of this figure of merit follows directly from the second law and corresponds to the Carnot limit, see Sec.~\ref{Sec_GTTherm} for details.
To further motivate this definition, it is instructive to consider the special case of Carnot-type engines, which operate between two fixed temperature levels $T_0$ and $T_1>T_0$.
For such cycles, the effective input becomes $U=\eta_\Carnot Q_1$, where $\eta_\Carnot \equiv 1-T_0/T_1$ denotes the Carnot factor and $Q_1$ is the heat uptake during the hot phase of the cycle, which is considered the input of a heat engine in classical thermodynamics.
The generalized efficiency \eqref{QT_Eff} thus reduces to the normalized thermal efficiency and we have $\eta = \eta_\thermal / \eta_\Carnot$ with $\eta_\thermal \equiv W/Q_1$. 
Hence, our approach is consistent with the standard model of 
classical thermodynamics.

In the plot in Fig.~\ref{Fig_Setup}(e), we observe that the efficiency $\eta$ approaches the ideal value $1$ only in the quasi-static limit $\T\rightarrow\infty$, where the power $P \equiv W/\T$ goes to zero.
At the same time, the power at fixed efficiency decreases with the tunneling energy $\Delta$.
In the following section, we will show that this behavior can be understood from general trade-off relations, which are determined by the single-jump statistics of the thermodynamic fluxes between the working system and its environment.

\section{General Theory} \label{Sec_GenTheor}

\subsection{Thermodynamics} \label{Sec_GTTherm}

For a general model of a quantum heat engine, we now consider an arbitrary multi-level system, whose Hamiltonian $H_t$ can be modulated through external fields to extract mechanical work.
The temperature $T_t$ of the reservoir that forms the environment is controlled by a heat source that provides thermal energy.
The thermodynamics of the system is governed by the first and the second law,%
\begin{equation}
	\dot{E}_t = \Phi_t - P_t
		\quad\text{and}\quad
	\Sigma_t = \dot{S}_t - \Phi_t/T_t \geq 0 .
\end{equation}
In the weak-coupling regime, which we focus on in this article, the internal energy and entropy of the system can be expressed in terms of its density matrix $\rho_t$ as \cite{AlickiJPhysA1979}%
\begin{equation} \label{GTTh_EnEnt}
	E_t = \tr{\rho_t H_t}
		\quad\text{and}\quad
	S_t = -\tr{\rho_t\ln[\rho_t]} .
\end{equation}
Furthermore, the rate of heat uptake from the environment and the extracted mechanical power are given by%
\begin{equation} \label{GTTh_HeatPow}
	\Phi_t = \tr{\dot{\rho}_t H_t}
		\quad\text{and}\quad
	P_t = - \tr{\rho_t \dot{H}_t} ,
\end{equation}
the symbol $\Sigma_t$ denotes the total rate of entropy production and dots indicate time derivatives.
Note that we set Boltzmann's constant to $1$ throughout.

Under continuous periodic driving, the system settles to a limit cycle state and its internal energy and entropy become periodic functions of time.
Integrating the first law over a full period $\T$ thus gives the identity%
\begin{equation} \label{GTTh_Output}
	Q = W ,
		\quad\text{with}\quad
	Q = \tint \Phi_t
		\;\;\text{and}\;\;
	W = \tint P_t
\end{equation}
being the mean heat uptake and work output per cycle.
Analogously, the second law leads to the relation%
\begin{align} \label{GTTh_EffInput}
	\Delta S_\tot &= (U-W)/T_0 \geq 0 
		\quad\text{with} \\[3pt]
	\Delta S_\tot &= \tint \Sigma_t
		\quad\text{and}\quad
	U = \tint \eta_t \Phi_t .
	\nonumber
\end{align}
Here, $\Delta S_\tot$ is the average total entropy production per cycle.
We further recall that $\eta_t \equiv 1-T_0/T_t$ is the instantaneous Carnot factor with respect to the base temperature $T_0 \leq T_t$ of the environment and that $U$ corresponds to the effective input of thermal energy from the heat source \cite{BrandnerPhysRevX2015, BrandnerPhysRevE2016}. 

The inequality \eqref{GTTh_EffInput} shows that the efficiency of a general engine cycle can be consistently defined as%
\begin{equation} \label{GTTh_Effy}
	\eta \equiv W/U \leq 1 .
\end{equation}
This figure attains its upper bound $1$ in the reversible limit, for which $\Delta S_\tot = 0$.
This condition, however, can be met in generic systems only under quasi-static driving leading to vanishing power.
A quantitative description of this trade-off between power and efficiency cannot be derived from the elementary laws of thermodynamics and requires a microscopic model for the dynamics of the working system, which we introduce in the next section. 

\subsection{Dynamical Model}
The density matrix of the working system evolves according to a linear master equation with the form \cite{Breuer2002}%
\begin{equation} \label{GTDM_ME}
	\dot{\rho}_t = \L_t[\rho_t]. 
\end{equation}
The structure of the generator $\L_t$ thereby depends on the coupling mechanism between the system and its environment and on the hierarchy of the involved time scales.
Here, we focus on the adiabatic weak-coupling regime, where the applied driving is slow and the system-environment interactions can be treated perturbatively.
That is, we assume that both the time scale of the driving and the thermalization time scale $\Gamma_{\text{th}}^{-1}$, where $\hbar\Gamma_{\text{th}}$ is the typical system-environment interaction energy, are long compared to both the unperturbed evolution of the working system and the relaxation dynamics of the surrounding reservoir \cite{AlickiJPhysA1979, AlbashNewJPhys2012, YamaguchiPhysRevE2017, DannPhysRevA2018}.
The fluctuations $\hat T_t$ of the reservoir temperature displayed in Fig.~\ref{Fig_Setup}(b) therefore do not have to be taken into account in the discussion of the system dynamics.
The generator $\L_t$ can then be divided into a unitary part, which describes the evolution of the bare working system, and a dissipation superoperator, which accounts for the influence of the environment, 
that is,%
\begin{equation} \label{GTDM_WCGen}
	\L_t[\rho_t] = -\frac{i}{\hbar} \comm{H_t}{\rho_t} + \D_t[\rho_t] .
\end{equation}
Here, $\comm A B \equiv AB - BA$ denotes the commutator.
Owing to micro-reversibility, the dissipation superoperator can be further decomposed into independent Markovian dissipation channels. 
Specifically, we have%
\begin{align} \label{GTDM_DBGen}
	\D_t[\rho_t] &= \asum \D^{\alpha+}_t[\rho_t] + \D^{\alpha-}_t[\rho_t]
		\quad\text{with} \\[3pt]
	\D^{\alpha\pm}_t[\rho_t] &\equiv
			\frac{1}{2} \comm[\big]{J^{\alpha\pm}_t \rho_t}{(J^{\alpha\pm}_t)^\dagger}
	   	+	\frac{1}{2} \comm[\big]{J^{\alpha\pm}_t}{\rho_t (J^{\alpha\pm}_t)^\dagger} . \nonumber
\end{align}
The jump operators $J^{\alpha+}_t$ and $J^{\alpha-}_t$, which respectively describe the emission and absorption of a photon with energy $\varepsilon^\alpha_t>0$ by the reservoir, fulfill the relation%
\begin{equation} \label{GTDM_EJO}
	\comm{H_t}{J^{\alpha\pm}_t} = \pm\varepsilon^\alpha_t J^{\alpha\pm}_t
\end{equation}
and the detailed-balance condition%
\begin{equation}\label{GTDM_DBJO} 
	(J^{\alpha -}_t)^\dagger = \exp[\varepsilon^\alpha_t/2T_t]\, J^{\alpha +}_t. 
\end{equation}
For a detailed discussion of the microscopic basis and the range of validity of the adiabatic weak-coupling approach, see for example Refs.~\cite{AlickiJPhysA1979, AlbashNewJPhys2012, YamaguchiPhysRevE2017, DannPhysRevA2018}.
To ensure that the working system settles to a unique limit cycle state, we further require that the jump operators connect all energy levels of the working system during a finite fraction of the cycle \footnote{
		Specifically, we assume that the Hilbert space of the working system has finite dimension and that the set of all jump operators is irreducible for a finite fraction of the cycle \cite{MenczelJPhysA2019}.
		Irreducibility here means that the commutant of the set of jump operators contains only multiples of the identity operator.
	}.\nocite{MenczelJPhysA2019}

Using the master equation \eqref{GTDM_ME} and the structure of the generator \eqref{GTDM_WCGen}, the rate of heat uptake and the total rate of entropy production can be expressed as~\cite{Binder2018}%
\begin{align}
	&\Phi_t = \tr{\D_t[\rho_t]H_t} 
		\;\;\text{and}\;\;
	\Sigma_t = \tr{\D_t[\rho_t]\bigl(\ln[R_t] {-} \ln[\rho_t]\bigr)}, \nonumber\\[3pt]
	&\text{where}\;\;
	R_t\equiv \exp[-H_t/T_t] \big/ \tr{\exp[-H_t/T_t]}
\label{GTDM_EntMic}
\end{align}
denotes the instantaneous Gibbs state of the working system.
The instantaneous Gibbs state is introduced here only for technical reasons and does in general not describe the actual state of the system.
Upon observing that $\D_t[R_t] = 0$ as a consequence of the conditions \eqref{GTDM_EJO} and \eqref{GTDM_DBJO}, it follows from Spohn's theorem that $\Sigma_t\geq 0$ for any $\rho_t$ \cite{SpohnJMathPhys1978}.
This result shows that the adiabatic weak-coupling approach is inherently consistent with the second law, for details, see Ref.~\cite{BrandnerPhysRevE2016}.

\subsection{Quantum Jump Statistics} \label{Sec_GTQJS}

To develop a quantum jump description of microscopic heat engines, we now assume that the reservoir can be continuously monitored such that the external observer obtains a channel-resolved quantum jump record $\R$ for every operation cycle of the device.
Extending the notation introduced in Sec.~\ref{Sec_QubitI}, we write%
\begin{equation}
	\R = \bigl\{(t_k,d_k,\alpha_k)\} ,
\end{equation}
where $t_k$ is the time at which the event $k$ in the dissipation channel $\alpha_k$ is detected and $d_k=\pm$ indicates whether a photon was emitted ($+$) or absorbed ($-$) by the reservoir.
After collecting sufficiently many records, the single-jump distribution%
\begin{equation} \label{GTQJS_Dist}
	\P[\hat{X}] = \frac{1}{\A} \mathbb{E}\Bigl[\ksum \delta\bigl[ \hat{X}-d_k X^{\alpha_k}_{t_k} \bigr] \Bigr]
\end{equation}
can be determined for every thermodynamic flux $\hat{X}$ that is exchanged between the system and the reservoir.
Recall that $\mathbb{E}$ denotes the average over all possible records and $\A$ is the mean number of jumps per cycle; by $X^\alpha_t$, we denote the amount of the quantity $\hat{X}$ that is carried by a photon in the channel $\alpha$ at the time $t$.
For example, the fluxes of heat and of effective thermal energy are characterized by $Q^\alpha_t = \varepsilon^\alpha_t$ and $U^\alpha_t = \varepsilon^\alpha_t \eta^\alpha_t$.
Note that positive and negative values of the single jump variable $\hat X$ correspond to emission and absorption events, respectively.

To derive an explicit expression for the distribution \eqref{GTQJS_Dist}, we have to analyze the dynamics of the engine under continuous monitoring.
To this end, we use the stochastic wave function method, which unravels the master equation \eqref{GTDM_ME} into measurement-conditioned quantum trajectories with piecewise deterministic evolution of the pure state $\ket{\psi_t}$
of the system \cite{DalibardPhysRevLett1992, MolmerJOptSocAmB1993, PlenioRevModPhys1998, Breuer2002, BreuerPhysRevA2003}.
In this approach, every detected event $(d,\alpha)$ corresponds to a quantum jump, which is described by the transformation%
\begin{equation}
	\ket{\psi_t} \rightarrow \ket{\psi'_t}
		= J^{\alpha d}_{t}\ket{\psi_t} \big/ \norm[\big]{ J^{\alpha d}_{t}\ket{\psi_t} } .
\end{equation}
Here, $\norm{\ket\psi}^2 \equiv \braket\psi\psi$ denotes the norm of the state $\ket\psi$.
Between two consecutive jumps at the times $t$ and $t' > t$, the state changes continuously according to the transformation%
\begin{equation} \label{GTQJS_JEv}
	\ket{\psi'_t} \rightarrow \ket{\psi_{t'}}
		= W_{t',t}\ket{\psi'_t}	\big/ \norm[\big]{ W_{t',t}\ket{\psi'_t} } ,
\end{equation}
where the non-unitary time evolution operator is given by the anti-chronologically ordered exponential%
\begin{equation} \label{GTQJS_ConEv}
	W_{t',t} \equiv \overleftarrow{\exp}\bigg[
		{-}\frac{i}{\hbar} \intd{t}{t'}{\tau} K_\tau
	\bigg]
\end{equation}
with the effective Hamiltonian%
\begin{equation} \label{GTQJS_EffH}
	K_t \equiv H_t - \frac{i\hbar}{2} \asum
		(J^{\alpha+}_t)^\dagger J^{\alpha+}_t + (J^{\alpha-}_t)^\dagger J^{\alpha-}_t. 
\end{equation}
Hence, if the record $\R$ is observed over the period $\T$, the initial state $\ket{\psi_0}$ undergoes the transformation%
\begin{equation}
	\ket{\psi_0} \rightarrow \ket{\psi_\T[\R]}
		= W[\R]\ket{\psi_0} \big/ \norm[\big]{ W[\R]\ket{\psi_0} } ,
\end{equation}
where the record-conditioned time evolution operator%
\begin{equation} \label{GTQJS_TotEv}
	W[\R] \equiv W_{\T,t_M}^{\phantom{\alpha_k}}
		\overleftarrow{\phantom{\prod}}\hspace*{-13pt}\prod\nolimits_{k=1}^M
		J^{\alpha_k d_k}_{t_k} W_{t_k,t_{k-1}}^{\phantom{\alpha_k}}
\end{equation}
is found by successively applying the transformation rules \eqref{GTQJS_JEv} and \eqref{GTQJS_ConEv}.
The arrow in Eq.~\eqref{GTQJS_TotEv} indicates the the product is ordered anti-chronologically, $M$ is the total number of events in the record $\R$ and we set $t_0 \equiv 0$.

The probability density to observe a given record $\R$ for the initial state $\ket{\psi_0}$ can now be expressed as%
\begin{equation} \label{eq:path_weights}
	p[\R|\psi_0] = \norm[\big]{ W[\R]\ket{\psi_0} }^2 .
\end{equation}
Consequently, if the system is initially in the mixed state $\rho_0 = \sum\nolimits_j r^j_0\ket{\psi^j_0}\bra{\psi^j_0}$, the cycle average of any record-dependent observable $\mathcal{X}$ can be expressed as%
\begin{equation}
	\mathbb{E}[\mathcal{X}] = \jsum r^j_0 \intD{0}{\T}{[\R]} \mathcal{X}[\R]\, p[\R|\psi^j_0] .
\end{equation}
Here, $\int_0^\T \mathcal{D}[\R]$ denotes the sum over all records between $0$ and $\T$ and the function $\mathcal X[\R]$ assigns the corresponding value of the observable $\mathcal X$ to a given record $\R$.
This formula makes it possible to evaluate the distribution \eqref{GTQJS_Dist} in terms of the weights \eqref{eq:path_weights}, which leads to the compact expression%
\begin{align} \label{GTQJS_Mom}
	\av{\hat{X}^n}
		&\equiv \indefInt{\hat{X}} \hat{X}^n\, \P[\hat{X}] \\
		&= \frac{1}{\A} \asum \tint \bigl(
				j^{\alpha+}_t (X^{\alpha}_t)^n + j^{\alpha-}_t (-X^\alpha_t)^n
			\bigr) \nonumber
\end{align}
for the moments of the single-jump distribution $\P[\hat{X}]$ as we show in App.~\ref{App_SJD}.
The variables%
\begin{equation} \label{GTQJS_PhFlux}
	j^{\alpha\pm}_t \equiv \tr{\rho_t\, (J^{\alpha\pm}_t)^\dagger J^{\alpha\pm}_t}
\end{equation}
correspond to the mean flux of photons that is absorbed $(+)$ or emitted $(-)$ by the system through the channel $\alpha$ at the time $t$ and the activity $\A$ is the mean total number of jumps per cycle,%
\begin{equation} \label{GTQJS_ActDef}
	\A = \asum \tint \bigl( j^{\alpha+}_t + j^{\alpha-}_t \bigr) .
\end{equation}

\subsection{Bounds on Entropy Production}

The total entropy production $\Delta S_\tot$ provides a measure for the thermodynamic cost of running a cyclic heat engine in finite time.
In the following, we first show that this cost can be divided into two non-negative contributions, one arising from quantum jumps and one stemming from the decay of coherences.
We then derive a lower bound on the jump entropy production, which depends only on the activity $\A$ and the dimensionless parameter%
\begin{equation} \label{GTBE_Hom}
	\lambda_{\hat{X}} \equiv \sqrt{\av{\hat{X}}^2 \big/ \av{\hat{X}^2}} \leq 1 ,
\end{equation}
which we refer to as the \emph{homogeneity} of the flux $\hat{X}$.
These results will provide the basis for the derivation of our new trade-off relation between power and efficiency.  

\subsubsection{Decomposition of entropy production} \label{Sec_GTBEPTT}

We begin our analysis by observing that, upon inserting the spectral decomposition of the density matrix, $\rho_t =\sum\nolimits_j r^j_t\ket{\psi^j_t}\bra{\psi^j_t}$, the expressions \eqref{GTDM_EntMic} and \eqref{GTQJS_PhFlux} for the total rate of entropy production and the average photon fluxes can be rewritten as%
\begin{align} \label{GTBEPtt_SigDecomp}
	\Sigma_t &= \asum \sum\nolimits_{j\ell} \Gamma_t^{\alpha,j\ell}	g\bigl[r^\ell_t,r^j_t\exp[\varepsilon^\alpha_t/T_t]\bigr] , \\[3pt]
	j^{\alpha+}_t &= \sum\nolimits_{j\ell} \Gamma^{\alpha,j\ell}_t r^\ell_t
		\quad\text{and} \nonumber \\[3pt]
	j^{\alpha-}_t &= \sum\nolimits_{j\ell} \Gamma^{\alpha,j\ell}_t r^j_t\, \exp[\varepsilon^\alpha_t/T_t] \nonumber
\end{align}
with $\Gamma^{\alpha,j\ell}_t \equiv \abs{ \bra{\psi^j_t} J^{\alpha+}_t \ket{\psi^\ell_t} }^2$ and $g[a,b] \equiv (a-b)\ln[a/b]$.
Since the function $g[a,b]$ is convex for $a,b \geq 0$, we can apply Jensen's inequality \cite{Hardy1952}, which yields \footnote{
	Recall that Jensen's inequality can be formulated as follows.
	For two sets of real numbers $\{\varphi_j\} \subset \mathbb{R}^+$ and $\{x_j\} \subset D \subseteq \mathbb{R}$, with $\phi\equiv\sum_j\varphi_j<\infty$, and a function $f$ that is convex on $D$, we have%
	\begin{equation*}
		\sum\nolimits_j \varphi_j f[x_j] \geq \phi\, f\Bigl[\sum\nolimits_j \frac{\varphi_j}{\phi} x_j \Bigr] .
	\end{equation*}
	The analogous relation%
	\begin{equation*}
		\int_D dx\; \varphi[x]f[x] \geq \phi\, f\biggl[ \int_D dx\; \frac{\varphi[x]}{\phi} f[x] \biggr]
	\end{equation*}
	holds for any non-negative function $\varphi[x]$ on $D$ with $\phi \equiv \int_D dx \; \varphi[x] <\infty$.}%
\begin{equation} \label{GTBEPtt_SBnd}
	\Sigma_t \geq \asum g\bigl[j^{\alpha+}_t,j^{\alpha-}_t\bigr] .
\end{equation}
After integrating both sides of this relation over a full cycle, we end up with the result%
\begin{equation} \label{GTBEPtt_DBnd}
	\Delta S_\tot \geq \Delta S_\jump
		\equiv \asum\tint \bigl( j^{\alpha+}_t {-} j^{\alpha-}_t \bigr) \ln\bigl[ j^{\alpha+}_t \big/ j^{\alpha-}_t\bigr] \geq 0.  
\end{equation}
This bound admits a transparent physical interpretation, which derives from the observation that the quantity $\Delta S_\jump$ can be expressed as%
\begin{equation}
	\Delta S_\jump = \A\, \av{\hat{\Sigma}_{{{\rm j}}}}
		\quad\text{with}\quad
	\Sigma^{\alpha}_{\jump t} \equiv \ln\bigl[j^{\alpha+}_t \big/ j^{\alpha-}_t\bigr] .
\end{equation}
In analogy to the entropy production associated with classical stochastic dynamics on a discrete set of states, which is given by the same formal expression \cite{SeifertRepProgPhys2012}, we identify the flux $\hat{\Sigma}_\jump$ as the entropy production of single quantum jumps.
The quantity $\Delta S_\jump$ thus provides a measure for the average thermodynamic cost of all jumps in one cycle.
The remainder of the total entropy production,%
\begin{equation}
	\Delta S_\quant \equiv \Delta S_\tot - \Delta S_\jump \geq 0 ,
\end{equation}
stems from the non-unitary evolution of the system between the jumps, that is, from the decay of superpositions between different energy levels \footnote{
	Note that a similar decomposition of the total entropy production into a jump and a drift part was introduced in Ref.~\cite{LeggioPhysRevA2013} to derive quantum corrections to the integral fluctuation theorem.}. \nocite{LeggioPhysRevA2013}
It can therefore be interpreted as a measure for the thermodynamic cost of coherence.
As we show in App.~\ref{App_QCL}, the contribution $\Delta S_\quant$ indeed vanishes in the quasi-classical regime, where the density matrix of the system commutes with its Hamiltonian throughout the cycle and every jump operator can be identified with a single transition between two energy levels.
Under these conditions, equality is attained in Eq.~\eqref{GTBEPtt_SBnd} and the bound \eqref{GTBEPtt_DBnd} becomes trivial.

\subsubsection{Homogeneity bound}\label{Sec_GTBEHom}
In order to derive a lower bound on the jump entropy production, we first introduce the weighting factors and the rescaled photon fluxes%
\begin{align}
	\Lambda^\alpha_t &\equiv (X^\alpha_t)^2 (j^{\alpha+}_t + j^{\alpha-}_t) / \A \geq 0
		\quad\text{and} \\[3pt]
	k^{\alpha\pm}_t &\equiv	\frac{2j^{\alpha\pm}_t}{X^\alpha_t(j^{\alpha+}_t + j^{\alpha-}_t)} , \nonumber
\end{align}
which fulfill the relations%
\begin{align}
	\asum\tint \Lambda^\alpha_t &= \av{\hat{X}^2}
		\quad\text{and} \\[3pt]
	\asum\tint \Lambda^\alpha_t k^{\alpha\pm}_t &=	%\av{\abs{ \hat{X} }}
													\Xabs \pm \av{\hat{X}} . \nonumber
\end{align}
We thereby defined the auxiliary variable%
\begin{equation}
	\Xabs \equiv \frac 1 \A \asum\tint X^\alpha_t(j^{\alpha+}_t + j^{\alpha-}_t) .
\end{equation}
The expression \eqref{GTBEPtt_DBnd} can now be cast into the form%
\begin{align}
	\Delta S_\jump = \frac{\A}{4} \asum\tint \Lambda^\alpha_t h[k^{\alpha+}_t, k^{\alpha-}_t]
\end{align}
with $h[a,b] \equiv (a^2-b^2)\ln[a/b]$. 
Since this function is convex for $a,b\geq 0$, Jensen's inequality implies \footnote{
	We recall that $\artanh[a] \equiv \frac{1}{2}\ln[\frac{1+a}{1-a}]$ for $-1<a<1$.}%
\begin{align} \label{GTBEhb_DBnd1}
	\Delta S_\jump
		&\geq \frac{\A}{4\av{\hat{X}^2}} h\bigl[	\Xabs + \av{\hat{X}},
													\Xabs - \av{\hat{X}}) \bigr] \\[3pt]
		&= 2\A\, \frac{\abs{\av{\hat X}}\; \abs{\Xabs}}{\av{\hat{X}^2}} \artanh\bigl[ \abs{\av{\hat{X}}} \big/ \abs{\Xabs} \bigr] . \nonumber
\end{align}
Finally, because the right-hand side of the inequality \eqref{GTBEhb_DBnd1} is monotonically decreasing in $\abs{\Xabs}$, this variable can be eliminated by replacing it with its upper bound%
\begin{equation}
	\A\sqrt{\av{\hat{X}^2}} \geq \abs{\Xabs} ,
\end{equation}
which again follows from Jensen's inequality.
Recalling the definition \eqref{GTBE_Hom} thus leaves us with the compact result%
\begin{equation} \label{GTBEhb_HB}
	\Delta S_\jump \geq 2\A\, \lambda_{\hat{X}} \artanh[\lambda_{\hat{X}}] ,
\end{equation}
which shows that the jump entropy production is boun\-ded from below by a monotonically increasing function of the homogeneity of any thermodynamic flux $\hat{X}$.

This figure attains its upper limit $1$, for which the right-hand side of Eq.~\eqref{GTBEhb_HB} diverges, if the corresponding single-jump distribution $\P[\hat{X}]$ has zero width, indicating that either emissions or absorptions are fully suppressed and every photon carries the same amount of the quantity $\hat{X}$.
Any deviation of $\lambda_{\hat{X}}$ from $1$ signifies fluctuations in the single-jump units of $\hat{X}$ with the lower limit $0$ being attained if no net exchange of the quantity $\hat{X}$ takes place between the system and the reservoir, i.e., if $\av{\hat{X}}=0$.
The relation \eqref{GTBEhb_HB} then reduces to the trivial bound $\Delta S_\jump \geq 0$.

\subsection{Performance Bounds for Quantum Heat Engines} \label{Sec_PBQHE}

Our bounds on entropy production \eqref{GTBEPtt_DBnd} and \eqref{GTBEhb_HB} imply a whole family of trade-off relations between power and efficiency, which we derive in two steps.
In the first one, we obtain a simple relation, which depends on the second single-jump moment of the effective thermal input and allows us to recover two earlier results.
We then derive an optimal trade-off relation, which is stronger than the simple one but involves more parameters.

\subsubsection{Simple trade-off relation}

We first consider the effective thermal input $\hat{U}$ and note that the bounds \eqref{GTBEPtt_DBnd} and \eqref{GTBEhb_HB} together imply%
\begin{equation}
	\Psi\Delta S_\tot \geq 2\A\, \lambda_{\hat{U}}\artanh[\lambda_{\hat{U}}]
		\;\; \text{with}\;\;
	\Psi \equiv\Delta S_\jump / \Delta S_\tot \leq 1 .
\end{equation}
Upon recalling Eqs.~\eqref{QT_Eff}, \eqref{GTTh_EffInput} and \eqref{GTBE_Hom}, the first of these bounds can be rewritten as a trade-off relation between the power and the efficiency, which is given by%
\begin{equation} \label{PBQHEMB_InPETOR}
	P \leq \eta \gamma \sqrt{\av{\hat{U}^2}} \tanh\left[ \frac{\Psi(1-\eta)}{2T_0}\sqrt{\av{\hat{U}^2}} \right]
\end{equation}
with $\gamma\equiv \A/\T$ denoting the average jump rate.
Note that we used $\av{\hat U} = U / \A = P / (\eta\gamma)$ in the derivation.
Our trade-off relation shows that, for generic systems with finite $\gamma$, the power output of any cyclic engine must go to zero as its efficiency approaches the ideal value $1$.
The linear slope of this decay is determined by the second single-jump moment of the thermal input $\hat{U}$.
Moreover, while all quantities in the bound \eqref{PBQHEMB_InPETOR} are generally interrelated, this bound restricts the possible values of the performance indicators $P$ and $\eta$ for any given values of the remaining parameters.
This restriction becomes successively stronger as the parameter $\Psi$ decreases, that is, as the coherence-induced entropy production $\Delta S_\quant = \Delta S_\tot - \Delta S_\jump$ increases.
In line with previous results \cite{KosloffEntropy2013, BrandnerPhysRevE2016, BrandnerPhysRevLett2017, BrandnerPhysRevLett2020}, this behavior indicates that coherence is generally detrimental to the performance of microscopic heat engines, at least under weak-coupling and slow-driving conditions.

The trade-off relation \eqref{PBQHEMB_InPETOR} includes two earlier results as special cases.
First, for small driving amplitudes, it reduces to the bound that was obtained in Ref.~\cite{BrandnerPhysRevE2016} as we show in App.~\ref{App_LRR}.
Second, for Carnot-type cycles with two heat baths at different temperatures, Eq.~\eqref{PBQHEMB_InPETOR} becomes%
\begin{align}
	P &\leq \eta_\thermal \gamma \sqrt{\av{\hat{Q}_1^2}} \tanh\left[ \frac{\Psi(\eta_\Carnot-\eta_\thermal)}{2T_0}
	\sqrt{\av{\hat{Q}^2_1}}\right]
		\quad\text{with} \nonumber \\[3pt]
	\av{\hat{Q}^2_1} &= \frac{1}{\A} \asum\intd{\T_1}{\hphantom{\T}}{t} (\varepsilon^\alpha_t)^2(j^{\alpha+}_t + j^{\alpha-}_t) \label{PBQHEMB_CarTOR}
\end{align}
denoting the second single-jump moment of the heat uptake during the hot phase of the cycle $\T_1$; recall that $\eta_\Carnot$ and $\eta_\thermal$ denote the Carnot factor and the thermal efficiency and that $\hat U = \eta_\Carnot \hat Q_1$.
Upon noting that $\tanh[a]\leq a$ for $a\geq 0$, this bound can be reduced to the weaker trade-off relation%
\begin{equation} \label{PBQHEMB_CarPETOR}
	P\leq \eta_\thermal (\eta_\Carnot - \eta_\thermal)\, \Theta / T_0
		\quad \text{with} \quad
	\Theta \equiv \gamma \av{\hat{Q}^2_1}/2,
\end{equation}
which was derived in Refs.~\cite{ShiraishiPhysRevLett2016, ShiraishiJStatPhys2019} by Shiraishi and co-workers.

\begin{figure*}
	\centering
	\includegraphics[width=\textwidth]{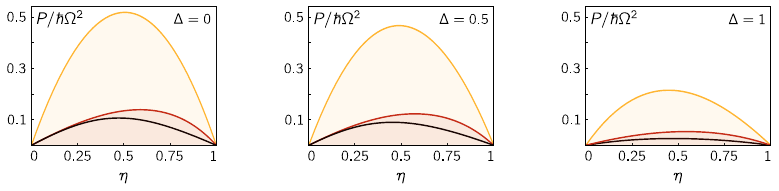}
	\caption{Power and efficiency of the qubit engine for three different tunneling energies.
		In each panel, the three curves, from top to bottom, show the simple bound \eqref{PBQHEMB_InPETOR}, the optimal bound \eqref{PBQHEOB_OptTOR} and the actual power output of the engine as a function of its efficiency.
		The shaded areas under the two upper curves indicate the admissible regions of the power-efficiency plane for the corresponding bounds.
		All plots were prepared by varying the cycle time from $\T=0.1/\Omega$ to $\T=250/\Omega$ for $\kappa=10$ and $T_0=\hbar\Omega$.}
	\label{Fig_2}
\end{figure*}

Before moving on, it is worth noting that applying the bound \eqref{GTBEhb_HB} to the heat uptake $\hat{Q}$ instead of the thermal input $\hat{U}$ yields the alternative trade-off relation%
\begin{equation}\label{PBQHEMB_OutPETOR}
	P\leq \eta\gamma \sqrt{\av{\hat{Q}^2}/\eta^2} \tanh\left[ \frac{\Psi(1-\eta)}{2T_0}\sqrt{\av{\hat{Q}^2}/\eta^2}\right] ,
\end{equation}
which is, however, weaker than the one in Eq.~\eqref{PBQHEMB_InPETOR}, since $\av{\hat{U}^2}\leq \av{\hat{Q}^2}\leq \av{\hat{Q}^2}/\eta^2$ and the hyperbolic tangent is a monotonically increasing function.

\subsubsection{Optimal trade-off relation}

We now consider the flux%
\begin{equation}
	\hat{Y}_{\varphi} \equiv \hat{U} + \varphi\hat{Q} ,
\end{equation}
whose first and second moment are given by%
\begin{align}
	\av{\hat{Y}_\varphi} &= \av{\hat{U}} + \varphi\av{\hat{Q}} = (U + \varphi W) / \A
		\quad\text{and} \\[3pt]
	\av{\hat{Y}_{\varphi}^2} &= \av{\hat{U}^2} + 2\varphi\av{\hat{R}^2} + \varphi^2\av{\hat{Q}^2} \nonumber
\end{align}
with $R^{\alpha}_t \equiv \varepsilon^\alpha_t\sqrt{\eta_t}$ and $\varphi$ being an arbitrary real number.
Upon applying the bound \eqref{GTBEhb_HB}, this ansatz yields the general trade-off relation%
\begin{align} \label{PBQHEOB_GenTOR}
	&P \leq \eta\gamma \sqrt{\av{\hat{Y}_\varphi^2} / (1 + \varphi\eta)^2} \\
	&\qquad\qquad \times \tanh\left[
		\frac{\Psi(1-\eta)}{2T_0} \sqrt{\av{\hat{Y}_\varphi^2}/(1 + \varphi\eta)^2}
	\right] , \nonumber
\end{align}
which includes the two results \eqref{PBQHEMB_InPETOR} and \eqref{PBQHEMB_OutPETOR} as limiting cases for $\varphi\to 0$ and $\varphi \to \infty$, respectively. 
Its strongest form is obtained by choosing $\varphi$ such that the right-hand side of the inequality \eqref{PBQHEOB_GenTOR} becomes minimal.
This value, which can be found by inspection, also maximizes the homogeneity $\lambda_{\hat{Y}_\varphi}$ of the flux $\hat{Y}_\varphi$ and is given by%
\begin{equation}
	\varphi^\ast = -\frac{ \av{\hat{R}^2} - \eta\av{\hat{U}^2} }{ \av{\hat{Q}^2} - \eta\av{\hat{R}^2} } .
\end{equation}
Inserting this result into Eq.~\eqref{PBQHEOB_GenTOR} gives the optimal trade-off relation%
\begin{align} \label{PBQHEOB_OptTOR}
	P &\leq \eta\gamma z_\eta \tanh\left[ \frac{\Psi(1-\eta)}{2T_0} z_\eta \right]
		\quad\text{with} \\[3pt]
	z^2_\eta &\equiv \frac{ \av{\hat{Q}^2}\av{\hat{U}^2} - \av{\hat{R}^2}^2 }{ \av{\hat{Q}^2} - 2\eta\av{\hat{R}^2} + \eta^2\av{\hat{U}^2} } . \nonumber
\end{align}
As we will show in the following section, this bound can be significantly stronger than the simple one in Eq.~\eqref{PBQHEMB_InPETOR}.
We stress that, despite its complex structure, the trade-off relation \eqref{PBQHEOB_OptTOR} could be tested in experiments since it involves only parameters that would be accessible through single-photon measurements.

\section{Qubit Engine: New Bounds} \label{Sec_QubitII}

To probe the quality of our new trade-off relations, we now return to the qubit engine discussed in Sec.~\ref{Sec_QubitI}. 
The dissipative dynamics of this system can be described with the two jump operators \cite{KupiainenPhysRevE2016}%
\begin{equation} \label{QT2_JOp}
	J^\pm = \sqrt{ \frac{\mp \kappa\Omega V}{1-\exp[\pm\hbar\Omega V/T]}} \ket{E^{\pm}} \bra{E^{\mp}},
\end{equation}
where the dimensionless parameter $\kappa$ determines the average jump frequency and $\ket{E^+}$ and $\ket{E^-}$ are the eigenvectors of the Hamiltonian \eqref{QT_Hamiltonian} with corresponding eigenvalues $E^\pm = \pm\hbar\Omega V$.
Upon inserting the protocols \eqref{QT_Prot} for the level splitting $V$ and the base temperature of the reservoir $T$, the periodic density matrix of the qubit can be determined by numerically solving the master equation \eqref{GTDM_ME}.
The work output $W$, the thermal input $U$ and the total entropy production $\Delta S_\tot$ of the engine can then be evaluated using Eqs.~\eqref{GTTh_HeatPow}, \eqref{GTTh_Output} and \eqref{GTTh_EffInput} \cite{BrandnerPhysRevLett2020}.
Furthermore, the second single-jump moments of the fluxes $\hat{Q}$, $\hat{U}$, $\hat{R}$ and the first moment of 
$\hat{\Sigma}_\jump$, which enter the trade-off relations \eqref{PBQHEMB_InPETOR} and \eqref{PBQHEOB_OptTOR}, can be evaluated with the help of Eqs.~\eqref{GTQJS_PhFlux} and
\eqref{GTQJS_Mom}. 

The results of this analysis are plotted in Fig.~\ref{Fig_2}.
They show that the simple trade-off relation \eqref{PBQHEMB_InPETOR} overestimates the power of the qubit engine by a factor between $4$ and $9$.
By contrast, the optimal bound \eqref{PBQHEOB_OptTOR} closely follows the exact power-efficiency curve and practically saturates for small $\eta$.
Exact saturation is, in fact, achieved for small driving amplitudes and optimal protocols as we show in App.~\ref{App_LRR}.
As a second key observation, we find that the power at fixed efficiency is uniformly suppressed along with its upper bounds in the tunneling energy $\Delta$.
This behavior can be understood by noting the engine is quasi-classical in the limit $\Delta\rightarrow 0$, where the eigenstates of the Hamiltonian \eqref{QT_Hamiltonian} become independent of the level separation.
As $\Delta$ deviates from $0$, the driving generates superpositions between the two energy levels of the system.
This effect leads to coherence-induced dissipation and thus reduces the performance of the engine.

\section{Concluding Perspectives} \label{Sec_DisPers}

Power and efficiency are arguably the two most important benchmarks for the performance of a heat engine. 
Quantitative bounds that make it possible to assess the trade-off between these two figures are key results of the theory of microscopic heat engines that has emerged over the last years.
This paper contributes to these ongoing developments in two ways.
On the conceptual side, our analysis shows that, within the adiabatic weak-coupling regime, a whole family of trade-off relations between power and efficiency can be derived in a technically simple and transparent manner.
These relations, which unify and extend previous results, were obtained only through the repeated application of Jensen's inequality.
As we show in App.~\ref{App_MR}, it is straightforward to generalize this technique for setups involving multiple reservoirs and other types of thermal devices such as microscopic refrigerators.
From a practical perspective, our approach delivers a clear physical interpretation of the additional parameters that determine the relationship between the power and the efficiency of microscopic heat engines. 
Inspired by current developments in the area of superconducting circuits, our theory provides a promising avenue towards practical tests of thermodynamic trade-off relations in future experiments, which could shed new light on the working mechanisms of microscopic thermal devices.

Turning to more general situations, we note that our results provide a valuable starting point for investigations of the effects of fast driving and strong coupling on the power and the efficiency of heat engines.
In these regimes, the performance of heat engines can generally be enhanced through coherence, see for example \cite{ThomasPhysRevE2018, AbiusoPhysRevA2019, NewmanPhysRevE2020, MukherjeeCommunPhys2020}.
In order to analyze this performance boost on the basis of our trade-off relations, further theoretical research generalizing the concept of single-jump distributions to the strong-coupling and fast-driving regimes will be necessary.
In addition, a fully realistic model of small-scale calorimetric measurements must account for imperfect photon detection, for instance due to background noise, as well as the finite size of the electronic reservoir and the back action of its temperature fluctuations on the working system.
Investigating how our bounds will be altered by these effects is an important subject for future work.
In paving the way for such studies, our paper contributes to the general goal of a unified and experimentally confirmed theory of thermodynamic trade-off relations for microscopic thermal devices.

\begin{acknowledgments}
We thank J.~P.~Pekola and K.~Saito for helpful comments.
CF acknowledges financial support from the Academy of Finland (Projects No.\ 308515 and No.\ 312299).
KB has received funding for the research presented in this article from the Academy of Finland (Contract No.\ 296073), the University of Nottingham through a Nottingham Research Fellowship and from UK Research and Innovation through a Future Leaders Fellowship (Grant Reference: MR/S034714/1).
Authors at Aalto University are affiliated with the Centre of Quantum Engineering.
\end{acknowledgments}

\appendix

\section{Multiple Reservoirs}\label{App_MR}

To keep the discussion in the main text simple, we focused on heat engines that operate with a single reservoir in the main text.
In the following, we show how our approach can be applied to setups with several reservoirs, which cover more general types of thermal devices.
As an application, we derive a family of trade-off relations between cooling power and efficiency for microscopic refrigerators.

\subsection{Thermodynamics and dynamical model}

For a multi-reservoir setup, the first and the second law read%
\begin{equation}
	\dot{E}_t = \rsum \Phi^\nu_t - P_t
		\;\;\text{and}\;\;
	\Sigma_t = \dot{S}_t - \rsum \Phi^\nu_t/T^\nu_t\geq 0 .
\end{equation}
Here, $T^\nu_t$ is the periodically modulated temperature of the reservoir $\nu$, which provides the working system with the heat current $\Phi^\nu_t$.
To derive microscopic expressions for these currents, we recall that each reservoir can be described with a separate dissipation superoperator in the weak-coupling regime \cite{SpohnAdvChemPhys1978}.
Hence, the generator $\L_t$, which enters the master equation \eqref{GTDM_ME}, has the form%
\begin{equation}
	\L_t[\rho_t] = -\frac{i}{\hbar}[H_t,\rho_t] + \rsum \D^\nu_t[\rho_t] .
\end{equation}
Owing to micro-reversibility, each of the superoperators $\D^\nu_t$ can be further decomposed into independent dissipation channels, i.e., we have%
\begin{align}
	\D^\nu_t[\rho_t] &= \asum \D^{\nu\alpha+}_t[\rho_t] + \D^{\nu\alpha-}_t[\rho_t]
		\quad\text{with}\quad\\[3pt]
	\D^{\nu\alpha\pm}_t[\rho_t] 
		&\equiv \frac{1}{2} \comm{J^{\nu\alpha\pm}_t\rho_t}{(J^{\nu\alpha\pm}_t)^\dagger}
			+ \frac{1}{2} \comm{J^{\nu\alpha\pm}_t}{\rho_t(J^{\nu\alpha\pm}_t)^\dagger} .
	\nonumber
\end{align}
The jump operators $J^{\nu\alpha+}_t$ and $J^{\nu\alpha-}_t$, which describe the exchange of photons with energy $\varepsilon^{\nu\alpha}_t>0$ between the working system and the reservoir $\nu$, obey the relation%
\begin{equation}
	\comm{H_t}{J^{\nu\alpha\pm}_t} = \pm \varepsilon^{\nu\alpha}_t J^{\nu\alpha\pm}_t 
\end{equation}
and the detailed balance condition%
\begin{equation}
	(J^{\nu\alpha -}_t)^\dagger = \exp[\varepsilon^{\nu\alpha}_t/2T^\nu_t]\, J^{\nu\alpha +}_t .
\end{equation}
Upon recalling the expressions \eqref{GTTh_EnEnt} for the internal energy and entropy of the working system, the rate of heat uptake from the reservoir $\nu$ can now be identified as%
\begin{equation}
	\Phi^\nu_t = \tr{\D^{\nu}_t[\rho_t] H_t} .
\end{equation}
Furthermore, the total rate of entropy production $\Sigma_t$ can be decomposed as $\Sigma_t = \sum\nolimits_\nu \Sigma^\nu_t$, where each component%
\begin{align} \label{AppMR_REntTot}
	\Sigma^\nu_t &\equiv \tr{\D^\nu_t[\rho_t]\bigl(	\ln[R^\nu_t]-\ln[\rho_t] \bigr)} \geq 0
		\quad\text{with} \\[3pt]
	R^\nu_t &\equiv \exp[-H_t/T^\nu_t] \big/ \tr{\exp[-H_t/T^\nu_t]} , \nonumber
\end{align}
is non-negative according to Spohn's theorem \cite{SpohnJMathPhys1978}.

\subsection{Quantum jump statistics}

We now assume that each reservoir is monitored by means of an ultra-sensitive thermometer.
For every operation cycle, we thus obtain a quantum jump record%
\begin{equation}
	\R = \bigl\{(t_k, d_k, \alpha_k, \nu_k)\bigr\}
\end{equation}
with the variable $\nu_k$ indicating the reservoir where the event $k$ was detected.
After collecting sufficiently many records, the single-jump distributions%
\begin{equation} \label{AppMR_SJDist}
	\P_\nu[\hat{X}] = \frac{1}{\A_\nu} \mathbb{E}\left[
		\ksum \delta_{\nu\nu_k} \delta[\hat{X} - d_k X^{\nu_k\alpha_k}_{t_k}]
	\right]
\end{equation}
can be determined, where $X^{\nu\alpha}_t$ is the amount of the quantity $\hat{X}$ that is exchanged with a single photon in the channel $\alpha$ at the time $t$ between the working system and the reservoir $\nu$.
As we show in App.~\ref{App_SJD}, the activity $\A_\nu$, which corresponds to the mean number of events per cycle in the reservoir $\nu$, and the moments of the distributions \eqref{AppMR_SJDist}
can be expressed as%
\begin{equation} \label{AppMR_ActDef}
	\A_\nu = \tint \bigl( j^{\nu\alpha+}_t + j^{\nu\alpha-}_t \bigr) ,
\end{equation}
and%
\begin{align} \label{AppMR_MomGen}
	\av{\hat{X}^n}_\nu 
		&\equiv \indefInt{\hat{X}} \hat{X}^n \P_\nu[\hat{X}] \\
		&= \frac{1}{\A_\nu} \asum \tint \bigl( j^{\nu\alpha+}_t (X^{\nu\alpha}_t)^n + j^{\nu\alpha-}_t (-X^{\nu\alpha}_t)^n \bigr) , \nonumber
\end{align}
where%
\begin{equation}
	j^{\nu\alpha\pm}_t \equiv \tr{\rho_t (J^{\nu\alpha\pm}_t)^\dagger J^{\nu\alpha\pm}_t}
\end{equation}
denotes the average flux of photons that is absorbed $(-)$ or emitted $(+)$ by the reservoir $\nu$ through the channel $\alpha$ at the time $t$.

\subsection{Bounds on entropy production}

In order to generalize our bounds on entropy production \eqref{GTBEPtt_DBnd} and \eqref{GTBEhb_HB} for setups with multiple reservoirs, we first observe that the expressions \eqref{AppMR_REntTot} and \eqref{AppMR_MomGen} for the components of the total rate of entropy production and the reservoir-conditioned single-jump moments have the same formal structure as their single-reservoir counterparts, cf.\ Eqs.~\eqref{GTDM_EntMic} and \eqref{GTQJS_Mom}. 
Therefore, the steps of Sec.~\ref{Sec_GTBEPTT} can be repeated to obtain the bounds%
\begin{equation} \label{AppMR_DBnd}
	\Delta S^\nu_\tot \geq \Delta S^\nu_\jump = \A_\nu \av{\hat\Sigma_\jump}_\nu ,
\end{equation}
where $\Sigma^{\nu\alpha}_t \equiv \ln[j^{\nu\alpha+}_t/j^{\nu\alpha-}_t]$, and%
\begin{align}
	\Delta S^\nu_\tot &\equiv \tint\Sigma^\nu_t \geq 0 \quad\text{and} \\[3pt]
	\Delta S^\nu_\jump &\equiv \asum\tint \bigl( j^{\nu\alpha+}_t {-} j^{\nu\alpha-}_t \bigr) \ln\bigl[ j^{\nu\alpha+}_t \big/ j^{\nu\alpha-}_t \bigr] \geq 0 \nonumber
\end{align}
correspond to the total and the jump entropy production due to the reservoir $\nu$.
Second, by following the lines of Sec.~\ref{Sec_GTBEHom}, it is now straightforward to derive the bounds%
\begin{equation} \label{AppMR_HB}
	\Delta S^\nu_\jump \geq 2\A_\nu\, \lambda^\nu_{\hat{X}} \artanh[\lambda^\nu_{\hat{X}}] ,
\end{equation}
which generalize our previous result \eqref{GTBEhb_HB} in terms of the reservoir-resolved homogeneities%
\begin{equation} \label{AppMR_Hom}
	\lambda^\nu_{\hat{X}} \equiv \sqrt{\av{\hat{X}}^2_\nu \big/ \av{\hat{X}^2}_\nu}	\leq 1 .
\end{equation}

\subsection{Performance bounds for quantum refrigerators}

As an application of our multi-reservoir bounds \eqref{AppMR_DBnd} and \eqref{AppMR_HB}, we will now derive a family of thermodynamic trade-off relations for cyclic micro-coolers.
Such devices use a periodically driven microscopic working system to transfer heat from a cold reservoir with temperature $T_0$ to a hot one with temperature $T_1 > T_0$ \cite{MenczelPhysRevB2019}.
Their thermodynamic output and input are given by the cooling power $P_\cool \equiv Q_0/\T$ and the absorbed mechanical power $P_\ssIn \equiv -W/\T = (Q_1-Q_0)/\T$, respectively;
their thermal efficiency, or coefficient of performance, is defined as \cite{Zemansky1997}%
\begin{equation} \label{AppMR_RefEff}
	\omega\equiv P_\cool / P_\ssIn
		= Q_0 / (Q_1-Q_0)
		\leq \omega_\Carnot
		\equiv T_0 / (T_1-T_0) .
\end{equation}
Here, $Q_0>0$ and $Q_1>Q_0$ are the average heat extraction from the cold reservoir and heat disposal to the hot reservoir.
The upper bound $\omega_\Carnot$ on $\omega$ follows from the second law%
\begin{equation}
	\Delta S_\tot = Q_1 / T_1 - Q_0 / T_0 \geq 0 ,
\end{equation}
and corresponds to the Carnot limit for refrigerators.
As for heat engines, this bound is generically only attainable in the quasi-static limit, where $P_\cool=0$. 

For a quantitative account of the trade-off between cooling power and thermal efficiency, we apply our bounds \eqref{AppMR_DBnd} and \eqref{AppMR_HB} to the heat flux $\hat{Q}$ with $Q^{\nu\alpha}_t = \varepsilon^{\nu\alpha}_t$.
The resulting relation,%
\begin{align} \label{AppMR_RHom}
	&\Psi\Delta S_\tot \geq 2 \rsum \A_\nu\lambda^\nu_{\hat{Q}}\artanh[\lambda^\nu_{\hat{Q}}]
		\quad\text{with} \\[3pt]
	&\Psi \equiv \Delta S_{{{\rm j}}}/\Delta S_\tot
		= \Bigl( \rsum\Delta S^\nu_\jump \Bigr)	\Big/ \Bigl( \rsum\Delta S^\nu_\tot \Bigr) \leq 1, \nonumber
\end{align}
cannot be solved for $P_\cool$ analytically. 
We may, however, still obtain explicit trade-off relations by dropping either the first or the second summand on the right-hand side of the inequality \eqref{AppMR_RHom}, both of which are non-negative.
This strategy yields%
\begin{equation} \label{AppMR_RefTOR1}
	P_\cool \leq \gamma_0 \sqrt{\av{\hat{Q}^2}_0} \tanh\left[
		\frac{\Psi}{2T_1} \frac{\omega_\Carnot-\omega}{\omega_\Carnot\omega} \sqrt{\av{\hat{Q}^2}_0}
	\right]
\end{equation}
and%
\begin{equation} \label{AppMR_RefTOR2}
	P_\cool \leq \frac{\gamma_1 \omega}{1+\omega} \sqrt{\av{\hat{Q}^2}_1} \tanh\left[
		\frac{\Psi}{2T_1} \frac{\omega_\Carnot-\omega}{\omega_\Carnot(1+\omega)} \sqrt{\av{\hat{Q}^2}_1}
	\right] ,
\end{equation}
where $\gamma_\nu\equiv \A_\nu/\T$ is the average rate of events in the reservoir $\nu$ and we have used that $\abs{\av{\hat{Q}}_\nu} = Q_\nu / \A_\nu$ for $\nu=0,1$.
Alternatively, we can simplify the bound \eqref{AppMR_RHom} by noting that%
\begin{equation}
	\artanh[\lambda^\nu_{\hat{Q}}] \geq \lambda^\nu_{\hat{Q}}, 
\end{equation} 
thus obtaining the trade-off relation%
\begin{equation}\label{AppMR_RefTOR3}
	P_\cool	\leq 
		\frac{\Psi(\omega_\Carnot-\omega)}{2T_1\omega_\Carnot}
		\frac{ \omega\gamma_0\gamma_1 \av{\hat{Q}^2}_0 \av{\hat{Q}^2}_1 }{ (1+\omega)^2\gamma_0 \av{\hat{Q}^2}_0 + \omega^2\gamma_1 \av{\hat{Q}^2}_1 } .
\end{equation}
All three of the bounds \eqref{AppMR_RefTOR1}, \eqref{AppMR_RefTOR2} and \eqref{AppMR_RefTOR3} show that, first, the Carnot limit $\omega_\Carnot$ can, for finite jump rates $\gamma_\nu$, be attained only at the price of vanishing cooling power.
Second, the maximum cooling power at given efficiency decreases uniformly with the coherence factor $\Psi$.
Hence, like microscopic heat engines, micro-coolers can be expected to perform best in the quasi-classical limit, as has been observed before for qubit-based devices \cite{KarimiPhysRevB2016, MenczelPhysRevB2019}.
Which of the bounds \eqref{AppMR_RefTOR1}, \eqref{AppMR_RefTOR2} and \eqref{AppMR_RefTOR3} is strongest, in general, depends on the specific setting.

\section{Single-Jump Moments} \label{App_SJD}

In this appendix, we show how the expression \eqref{AppMR_MomGen} for the moments of the reservoir-conditioned single-jump distributions can be derived within the quantum jump approach to open-system dynamics.
For setups with one reservoir, this result reduces to Eq.~\eqref{GTQJS_Mom}.

We recall Eq.~\eqref{AppMR_SJDist} and use it to express the moments of the single-jump distributions in terms of the distribution of quantum jump records,%
\begin{equation} \label{eq:def_moments}
	\av{\hat X^n}_\nu = \frac{1}{\A_\nu} \intD{0}{\T}{[\R]} p[\R | \psi_0]\, \sum\nolimits_{k=1}^M\! \delta_{\nu\nu_k} \bigl( d_k X^{\nu_k\alpha_k}_{t_k} \bigr)^n .
\end{equation}
Here, $M$ is the number of events in the record $\R$ and we assumed, for simplicity, that the initial state of the system is the pure state $\ket{\psi_0}$.
To generalize Eqs.~\eqref{GTQJS_EffH} and \eqref{GTQJS_TotEv} for multiple-reservoir setups, we also defined the probability distribution of jump records,%
\begin{equation}
	p[\R|\psi_0]= \norm[\big]{ W[\R] \ket{\psi_0} }^2 ,
\end{equation}
where the record-conditioned time evolution operator is given by \cite{Breuer2002}%
\begin{align} \label{eq:appB_definitions}
	W[\R] &\equiv W_{\T,t_M} \overleftarrow{\phantom{\prod}}\hspace*{-13pt}\prod\nolimits_{k=1}^M J^{\nu_k\alpha_k d_k}_{t_k} W_{t_k,t_{k-1}}
		\quad\text{with} \\[3pt]
	W_{t',t} &\equiv \overleftarrow{\exp}\bigg[
		{-}\frac{i}{\hbar} \intd{t}{t'}{\tau} K_\tau
	\bigg]
		\quad\text{and}	\nonumber \\[3pt]
	K_t &\equiv H_t -\frac{i\hbar}{2} \sum\nolimits_{\nu,\alpha}\! (J^{\nu\alpha+}_t)^\dagger J^{\nu\alpha+}_t + (J^{\nu\alpha-}_t)^\dagger J^{\nu\alpha-}_t . \nonumber
\end{align}
The path integral $\int_0^\T \mathcal D[\R]$ is defined as the sum over all possible records,%
\begin{equation}
	\intD{0}{\T}{[\R]} \mathcal X[\R] \equiv \sum_M  \intd{0}{\T}{t_M} \smashoperator{\sum_{\nu_M\alpha_Md_M}} \;\;\cdots\;\; \intd{0}{t_2}{t_1} \smashoperator{\sum_{\nu_1\alpha_1d_1}} \mathcal X[\R] ,
\end{equation}
where $\mathcal X$ is any record-dependent observable.

In order to evaluate Eq.~\eqref{eq:def_moments}, we formally understand its right hand side as a function of the parameter $\T$ and derive a differential equation for $f(\T) \equiv \A_\nu \av{\hat X^n}_\nu$.
Writing out the path integral, $f(\T)$ can be expressed as
\begin{align}
	f(\T) &= \sum_M \intd{0}{\T}{t_M} \smashoperator{\sum_{\nu_M\alpha_Md_M}}
		\;\;\cdots\;\;
		\intd{0}{t_2}{t_1} \smashoperator[l]{\sum_{\nu_1\alpha_1d_1}} \\
	&\quad\;\,\times
		\bra{\psi_0} W[\R]^\dagger W[\R] \ket{\psi_0}
		\sum\nolimits_{k=1}^M\! \delta_{\nu\nu_k} \bigl( d_k X^{\nu_k\alpha_k}_{t_k} \bigr)^n . \nonumber
\end{align}
To determine the derivative of $f(\T)$, we will make use of the identities%
\begin{equation} \label{eq:appB_id1}
	\partial_\T W[\R] = -\frac{i}{\hbar} K_\T W[\R]
\end{equation}
and%
\begin{equation} \label{eq:appB_id2}
	W[\R] \big|_{t_M = \T} = J^{\nu_M\alpha_Md_M}_\T W[\R'] ,
\end{equation}
which follow directly from Eq.~\eqref{eq:appB_definitions}.
Here, the record $\R'$ is the record $\R$ without its last jump.
By combining Eqs.~\eqref{eq:appB_definitions} and \eqref{eq:appB_id2}, we further obtain the relation%
\begin{equation} \label{eq:appB_id2b}
	\smashoperator[r]{\sum_{\nu_M\alpha_Md_M}} W[\R]^\dagger W[\R] \big|_{t_M = \T}
		= \frac{i}{\hbar} W[\R']^\dagger (K_\T - K_\T^\dagger) W[\R'] .
\end{equation}

Recall that according to the Leibniz integral rule, the derivative of a parameter-dependent integral is generally given by
\begin{equation} \label{eq:appB_leibniz}
	\partial_\T \intd{0}{\T}{t} g(t, \T) = g(t, \T) \big|_{t = \T} + \intd{0}{\T}{t} \partial_\T\, g(t, \T)
\end{equation}
for any function $g$.
Using Eqs.~\eqref{eq:appB_id1}, \eqref{eq:appB_id2b} and \eqref{eq:appB_leibniz}, we find%
\begin{align}
	f'(\T) &= \intD{0}{\T}{[\R']} \smashoperator{\sum_{\nu_M\alpha_Md_M}} \delta_{\nu\nu_M} \bigl( d_M X^{\nu_M\alpha_M}_\T \bigr)^n \\[3pt]
	&\qquad \times \bra{\psi_\T[\R']} (J^{\nu_M\alpha_Md_M}_\T)^\dagger J^{\nu_M\alpha_Md_M}_\T \ket{\psi_\T[\R']} \nonumber \\[3pt]
	&= \asum \bigl( j^{\nu\alpha+}_\T (X^{\nu\alpha}_\T)^n + j^{\nu\alpha-}_\T (-X^{\nu\alpha}_\T)^n \bigr) . \nonumber
\end{align}
Recall that $\ket{\psi_\T[\R']}$ was defined as $W[\R']\ket{\psi_0}$.
Since%
\begin{equation}
	\av{\hat X^n}_\nu = \A_\nu^{-1} \tint f^{\prime}(t) ,
\end{equation}
our proof of Eq.~\eqref{AppMR_MomGen} is concluded.
The formula \eqref{AppMR_ActDef}, and thus Eq.~\eqref{GTQJS_ActDef}, for the ac\-tivity follows from the normalization condition $\av{1}_\nu=1$.

\section{Quasi-Classical Limit} \label{App_QCL}

In the adiabatic weak-coupling regime, coherence can enter a thermodynamic engine cycle in two different ways:
	through driving-induced superpositions between the energy levels of the working system and through superpositions between energetically degenerate transitions, which belong to the same dissipation channel.
A microscopic heat engine can thus be regarded as quasi-classical if the two conditions%
\begin{subequations}
\begin{align}
	&	\comm{H_t}{\rho_t} = 0 \quad\text{and} \label{AppQCL_GenCond1} \\[3pt]
	&	\tr{\bigl(J^{\alpha\pm}_t (J^{\alpha\pm}_t)^\dagger\bigr)^2}
			= \bigl(\tr{J^{\alpha\pm}_t (J^{\alpha\pm}_t)^\dagger}\bigr)^2 \label{AppQCL_GenCond2}
\end{align}
\end{subequations}
are met throughout the cycle.
In the following, we will explain the motivation for these conditions in more detail and show that they are sufficient and necessary for $\Delta S_\quant$ to vanish.
Note that we focus on settings with a single reservoir in this appendix for the sake of simplicity.

The first condition \eqref{AppQCL_GenCond1} states that there is no coherence in the working fluid at any time.
For example, this condition is satisfied at long times if the system Hamiltonians at any two times $t$ and $t'$ commute, $\comm{H_t}{H_{t'}} = 0$.
It ensures that the Hamiltonian and the periodic state of the working system share a common eigenbasis, i.e., that%
\begin{equation}
	H_t = \sum\nolimits_n\! E^n_t \ket{n_t}\bra{n_t}
		\quad\text{and}\quad
	\rho_t = \sum\nolimits_n\! r^n_t \ket{n_t}\bra{n_t} .
\end{equation}
Here, $E^1_t \leq E^2_t \leq \cdots$ are the ordered energy levels of the system and $\{\ket{n_t}\}$ is a complete set of orthogonal vectors at each time.

The second condition \eqref{AppQCL_GenCond2} implies, together with the conditions \eqref{GTDM_EJO} and \eqref{GTDM_DBJO}, that the jump operators have the form%
\begin{align}
	J^{\alpha+}_t &= w^{\alpha}_t \ket{m^{\alpha}_t}\bra{n^{\alpha}_t}
		\quad\text{and} \\[3pt]
	J^{\alpha-}_t &= \exp[-\varepsilon^\alpha_t/2T_t]\, (w^{\alpha}_t)^\ast \ket{n^{\alpha}_t}\bra{m^{\alpha}_t}
	\nonumber
\end{align}
with ${m^\alpha_t > n^\alpha_t}$ and the complex weighting factors $w^{\alpha}_t$ depending on the specifics of the system.
This condition thus ensures that the set $\{ j^{\alpha\pm}_t \}$ of directed photon currents in our model corresponds one-to-one to the set of directed probability currents in the classical thermal machine.
To illustrate the necessity of this condition, we examine the difference between two dynamical models for a three-level system with energy eigenstates $\ket{1}$, $\ket{2}$ and $\ket{3}$.
In the first model, the influence of the environment is modeled using two dissipation channels with jump operators $J^{12+} \equiv \ket{2}\bra{1}$ and $J^{23+} \equiv \ket{3}\bra{2}$.
In the second model, there is only a single jump operator $J^+ \equiv J^{12+} + J^{23+}$ and the corresponding directed photon current is given by $j^+_t = j^{12+}_t + j^{23+}_t$.
%If condition \eqref{AppQCL_GenCond1} is satisfied, both models induce the same time evolution of the system.
The latter model describes a setup where the two transitions are in superposition and the transition that took place cannot be inferred from an emitted photon.
A classical setup, where all types of state transition are in principle distinguishable, must therefore be described with a model of the first type.
Note that both models lead to the same time evolution of the system if condition \eqref{AppQCL_GenCond1} holds, but the jump entropy production agrees with the classical expression
\begin{equation}
	\Delta S_{\text{cl}} \equiv \tint \,\sum_{\mathclap{\alpha \in \{ 12, 23 \}}}\; \bigl( j^{\alpha+}_t {-} j^{\alpha-}_t \bigr) \ln\bigl[ j^{\alpha+}_t \big/ j^{\alpha-}_t\bigr]
\end{equation}
only in the first model.

If both conditions are satisfied, the coefficients $\Gamma^{\alpha,j\ell}_t$ introduced in Eq.~\eqref{GTBEPtt_SigDecomp} become%
\begin{equation} \label{AppQCL_GenCond3}
	\Gamma^{\alpha,j\ell}_t
		\equiv \abs{ \bra{j_t} J^{\alpha+}_t \ket{\ell_t} }^2
		= \abs{ w^{\alpha}_t }^2\, \delta_{jm^\alpha}\delta_{\ell n^\alpha},
\end{equation}
that is, they vanish for all but one combination of indices $j$ and $l$.
This form of the coefficients implies that equality is attained in Eq.~\eqref{GTBEPtt_SBnd}.
Conversely, whenever equality is attained in Eq.~\eqref{GTBEPtt_SBnd}, the coefficients must have this form and, hence, the jump operators must be given by $J^{\alpha\pm}_t \sim \ket{\psi^{\alpha\pm}_t}\bra{\varphi^{\alpha\pm}_t}$ for some states $\ket{\psi^{\alpha\pm}_t}$, $\ket{\varphi^{\alpha\pm}_t}$.
Due to the detailed-balance condition \eqref{GTDM_EJO}, these states must be eigenstates of the Hamiltonian and condition \eqref{AppQCL_GenCond2} is therefore satisfied.
From the form \eqref{AppQCL_GenCond3} of the coefficients $\Gamma^{\alpha,j\ell}_t$ and our assumption that the jump operators connect all energy levels of the working system, we finally deduce that the eigenstates of $\rho_t$ are also eigenstates of $H_t$ and condition \eqref{AppQCL_GenCond1} is satisfied as well.

We have thus shown that the two conditions \eqref{AppQCL_GenCond1} and \eqref{AppQCL_GenCond2} are both sufficient and necessary for equality in Eq.~\eqref{GTBEPtt_SBnd}.
It follows that we have $\Delta S_\tot = \Delta S_\jump$ in the quasi-classical limit and $\Delta S_\tot > \Delta S_\jump$ otherwise.
This result confirms that the contribution $\Delta S_\quant$ to the total entropy production is of genuine quantum origin and can be regarded as a measure for the thermodynamic cost of coherence.

\section{Linear-Response Regime} \label{App_LRR}

The physical picture behind our thermodynamic trade-off relations becomes particularly clear in the linear-response regime, as we will demonstrate in this appendix.
To keep our analysis as simple as possible, we focus on the quasi-classical limit and setups with a single reservoir.

We assume that the Hamiltonian of the working system and the temperature of the reservoir are given by%
\begin{equation}
	H_t = H_0 + \Delta_w G^w f^w_t 
		\quad\text{and}\quad
	T_t = T_0(1+\Delta_u f^u_t) ,
\end{equation}
where the operator $G^w$ corresponds to the degree of freedom that couples to the mechanical driving, $f^w_t$ and $f^u_t$ are dimensionless periodic functions and $\Delta_w, \Delta_u\ll 1$ are dimensionless parameters that control the strength of the time dependent perturbations.
To the lowest order in $\Delta_w$ and $\Delta_u$, the effective input and the work output of the engine are given by  \cite{BrandnerPhysRevE2016}%
\begin{align}
	U &= L_{uw} \Delta_u\Delta_w + L_{uu} \Delta_u^2
		\quad\text{and} \\[3pt]
	W &= L_{ww} \Delta_w^2 + L_{wu} \Delta_w\Delta_u . \nonumber
\end{align}
Here, the generalized kinetic coefficients are defined as%
\begin{equation} \label{AppLRR_KinCo}
	L_{xy} \equiv \tint\!\! \intd{0}{\infty}{\tau} \dot{C}^{xy}_\tau \dot{f}^x_t f^y_{t-\tau}
\end{equation}
for $x=u,w$ and $y=u,w$. 
Further,%
\begin{align} \label{AppLRR_Kubo}
	& C^{xy}_t \equiv {\corr{G^x}{G^y}}_t \\[3pt]
	&\equiv \frac{1}{T_0} \intd{0}{1}{\lambda} \Bigl( \tr{\tilde{G}^{x}_t R^\lambda_0 G^{y} R_0^{1-\lambda}} {-} \tr{\tilde{G}^{x}_t R_0} \tr{G^{y} R_0} \Bigr) \nonumber
\end{align}
denotes the Kubo correlation function with respect to the Gibbs state $R_0 \equiv\exp[-H_0/T_0] \big/ \tr{\exp[-H_0/T_0]}$ of the unperturbed system \cite{Kubo1991}.
Tildes in Eq.~\eqref{AppLRR_Kubo} indicate Heisenberg-picture operators, which evolve according to the equilibrium master equation%
\begin{equation}
	\dot{\tilde{O}}_t = \frac{i}{\hbar}[H_0,\tilde{O}_t] + \mathsf{F}_0[\tilde{O}_t]
		\quad\text{with}\quad
	\tilde{O}_{t=0} = O .
\end{equation}
The adjoint dissipation superoperator is defined as%
\begin{align}
	\mathsf{F}_0[O] &\equiv \asum \mathsf{F}^{\alpha+}_0[O] + \mathsf{F}^{\alpha-}_0[O] 
		\quad\text{with} \\[3pt]
	\mathsf{F}^{\alpha\pm}_0[O] &\equiv \frac{1}{2}(J_0^{\alpha\pm})^\dagger \comm{O}{J^{\alpha\pm}_0} + \frac{1}{2}\comm{(J^{\alpha\pm}_0)^\dagger}{O} J^{\alpha\pm}_0 \nonumber
\end{align}
and $J^{\alpha\pm}_0\equiv J^{\alpha\pm}_t\bigl|_{\Delta_w,\Delta_u=0}$ are the equilibrium jump operators.
Note that, in Eq.~\eqref{AppLRR_Kubo}, we have implicitly introduced the variable $G^u \equiv -H_0$ for convenience.
The subscript $0$ indicates equilibrium quantities throughout.

To derive the linear-response counterparts of our trade-off relations from Sec.~\ref{Sec_PBQHE}, we first observe that the bound \eqref{GTBEhb_HB} reduces to%
\begin{equation} \label{AppLRR_EBnd}
	\Delta S_\tot \geq 2 \A_0 \av{\hat{X}}^2\bigl/\av{\hat{X}^2}_0,
\end{equation}
since $\Delta S_\tot$ is of second order in the perturbations and the mean value of any thermodynamic flux must be of first order.
We now consider the flux $\hat{Y}'_\varphi \equiv (\hat{U}+\varphi\hat{Q})/\Delta_u$, whose first and second single-jump moments become%
\begin{align} \label{AppLRR_MomGenFlux}
	\av{\hat{Y}'_\varphi}
		&= U/(\A_0 \Delta_u) + \mathcal{O}[\Delta^2]\\[3pt]
		&= (L_{uw} \Delta_w + L_{uu}\Delta_u)/\A_0 +\mathcal{O}[\Delta^2]
	\quad\text{and}\nonumber\\
\av{\hat{Y}^{\prime 2}_\varphi}_0 &= \frac{1}{\A_0} \asum (\varepsilon^\alpha_0)^2 (j^{\alpha+}_0 + j^{\alpha-}_0) \tint (f^u_t + \varphi/\Delta_u)^2 \nonumber
\end{align}
in lowest order, since the parameter $\varphi$ must be considered as first order in $\Delta_u$ for consistency.
Inserting the expressions \eqref{AppLRR_MomGenFlux} into Eq.~\eqref{AppLRR_EBnd} yields the relation%
\begin{equation} \label{AppLRR_TORGen}
	P \leq \eta(1-\eta) \Theta_{\varphi,0}\Delta_u^2 / T_0
\end{equation}
with%
\begin{equation}
	\Theta_{\varphi,0}
		= \gamma_0 \av{\hat{Y}^{\prime 2}_\varphi}_0\bigl/2 
		= \frac{\vartheta_0}{\T} \tint (f^u_t + \varphi/\Delta_u)^2
\end{equation}
and%
\begin{align}
	\vartheta_0
		&\equiv \frac{1}{2}\asum (\varepsilon^\alpha_0)^2 (j^{\alpha+}_0 + j^{\alpha-}_0) \\[3pt]
		&= \frac{1}{2}\asum \Bigl(
				\tr{R_0\comm{H_0}{J^{\alpha+}_0}^\dagger \comm{H_0}{J^{\alpha+}_0}} \nonumber \\
		&\hspace*{3cm}
				+ \tr{R_0\comm{H_0}{J^{\alpha-}_0}^\dagger\comm{H_0}{J^{\alpha-}_0}}
			\Bigr) \nonumber \\[3pt]
		&= -\tr{R_0 H_0 \mathsf{F}_0[H_0]} . \nonumber
\end{align}
Here, we have used the conditions \eqref{GTDM_EJO} and \eqref{GTDM_DBJO}.

Setting $\varphi=0$ in Eq.~\eqref{AppLRR_TORGen} leads to the simple trade-off relation%
\begin{equation} \label{AppLRR_TORIn}
	P \leq \eta(1-\eta) \frac{\vartheta_0\Delta_u^2}{T_0} \frac{1}{\T}\tint (f^u_t)^2,
\end{equation}
which corresponds to Eq.~\eqref{PBQHEMB_InPETOR} for $\Psi=1$, since we focus on the quasi-classical limit here.
This result was derived earlier in Ref.~\cite{BrandnerPhysRevE2016}.
For two-temperature cycles, it becomes the linear-response version of the bound \eqref{PBQHEMB_CarPETOR}, which goes back to Refs.~\cite{ShiraishiPhysRevLett2016, ShiraishiJStatPhys2019}.

The bound \eqref{AppLRR_TORGen} becomes strongest for%
\begin{equation}
	\varphi^\ast = -\frac{\Delta_u}{\T} \tint f^u_t \equiv -\Delta_u \bar{f}^u ,
\end{equation}
as can be easily verified by inspection.
For this choice, we obtain the optimal trade-off relation%
\begin{equation} \label{AppLRR_TOROpt}
	P \leq \eta(1-\eta) \frac{\vartheta_0\Delta_u^2}{T_0} \frac{1}{\T} \tint (f^u_t -\bar{f}^u)^2 ,
\end{equation}
which corresponds to Eq.~\eqref{PBQHEOB_OptTOR}.
Note that the scaling factor between power and efficiency in the bounds \eqref{AppLRR_TORIn} and \eqref{AppLRR_TOROpt} is independent of the mechanical protocol $f^w_t$. 
Furthermore, Eq.~\eqref{AppLRR_TOROpt} implies the efficiency-independent bound
\begin{equation}
	P \leq \frac{\vartheta_0\Delta_u^2}{4T_0} \frac{1}{\T} \tint (f^u_t - \bar{f}^u)^2 ,
\end{equation}
on power, which was derived in Ref.~\cite{BrandnerPhysRevLett2017}.

Finally, it is instructive to note that the optimal trade-off relation \eqref{AppLRR_TOROpt} can be saturated if the variable $G^w$ is proportional to the unperturbed Hamiltonian $H_0$ and the equilibrium energy correlation function decays mono-exponentially, that is, if%
\begin{align} \label{AppLRR_OptCond}
	&	G^w = H_0/\zeta \quad\text{and} \\
	&	{\corr{H_0}{H_0}}_t = \Bigl( \tr{R_0(H_0)^2} - \bigl(\tr{R_0H_0}\bigr)^2 \Bigr) \exp[-\mu t] , \nonumber
\end{align}
where $\zeta$ and $\mu>0$ are real constants.
These conditions are met, for example, for the qubit engine discussed in the main text.
In this case, the mechanical protocol $f^{w\ast}_t$ that generates the maximal work for a given temperature protocol $f^u_t$ and fixed efficiency $\eta$ is given by%
\begin{equation} \label{AppLRR_OptProt}
	f^{w\ast}_t = \frac{\zeta\Delta_u}{\Delta_w} \Bigl( \eta f^u_t - \mu(1-\eta) \intd{0}{t}{\tau} (f^u_t - \bar{f}^u) \Bigr) .
\end{equation}
This result can be derived by expanding the protocols $f^w_t$ and $f^u_t$ into Fourier series and maximizing the work $W$ with respect to the Fourier coefficients of $f^w_t$ under the constraint $W - \eta U = 0$, for details see Refs.~\cite{BrandnerPhysRevE2016, BauerPhysRevE2016}.
Evaluating the kinetic coefficients \eqref{AppLRR_KinCo} for the protocols \eqref{AppLRR_OptProt} and using the conditions \eqref{AppLRR_OptCond} shows that the bound \eqref{AppLRR_TOROpt} is indeed saturated with%
\begin{align}
	\vartheta_0
		&= -\tr{R_0H_0\mathsf{F}_0[H_0]} \\
		&= \mu\Bigl( \tr{R_0(H_0)^2} - \bigl( \tr{R_0H_0} \bigr)^2 \Bigr) \nonumber
\end{align}
being proportional to the equilibrium energy fluctuations of the working system. 
Hence, we can conclude that our optimal trade-off relation between power and efficiency, Eq.~\eqref{PBQHEOB_OptTOR}, can be saturated in linear response.

\vbadness=10000
\hbadness=10000

\end{document}